\documentclass{SciPost}

\hypersetup{
    colorlinks,
    linkcolor={red!50!black},
    citecolor={blue!50!black},
    urlcolor={blue!80!black}
}

\usepackage[bitstream-charter]{mathdesign}
\DeclareSymbolFont{usualmathcal}{OMS}{cmsy}{m}{n}
\DeclareSymbolFontAlphabet{\mathcal}{usualmathcal}

\fancypagestyle{SPstyle}{
\fancyhf{}
\lhead{\colorbox{scipostblue}{\bf \color{white} ~SciPost Physics Proceedings }}
\rhead{{\bf \color{scipostdeepblue} ~Submission }}

\fancyfoot[C]{\textbf{\thepage}}
}

\usepackage{overpic}
\usepackage{tikz}
\usepackage{xcolor}
\usepackage{xspace}
\usepackage{graphicx}

\usepackage{newunicodechar}
\newunicodechar{★}{\star}

\definecolor{RRed}{RGB}{238,51,17}
\definecolor{RBlue}{RGB}{51,102,255}
\definecolor{RGreen}{RGB}{16,150,24}
\definecolor{ROrange}{RGB}{255, 153, 0}

\newcommand{\rivet}{\textsc{Rivet}\xspace}
\newcommand{\pythia}{\textsc{Pythia}\xspace}
\newcommand{\corsika}{\textsc{Corsika}\xspace}

\begin{document}

\pagestyle{SPstyle}

\begin{center}{\Large \textbf{\color{scipostdeepblue}{
\pythia 8 and Air Shower Simulations: A Tuning Perspective\\
}}}\end{center}

\begin{center}\textbf{
Chloé Gaudu\textsuperscript{1$\star$}
}\end{center}

\begin{center}
{\bf 1} Bergische Universität Wuppertal, Gaußstraße 20, 42119 Wuppertal, Germany
\\[\baselineskip]
$\star$ \href{mailto:email1}{\small gaudu@uni-wuppertal.de}
\end{center}

\definecolor{palegray}{gray}{0.95}
\begin{center}
\colorbox{palegray}{
  \begin{tabular}{rr}
  \begin{minipage}{0.36\textwidth}
    \includegraphics[width=55mm]{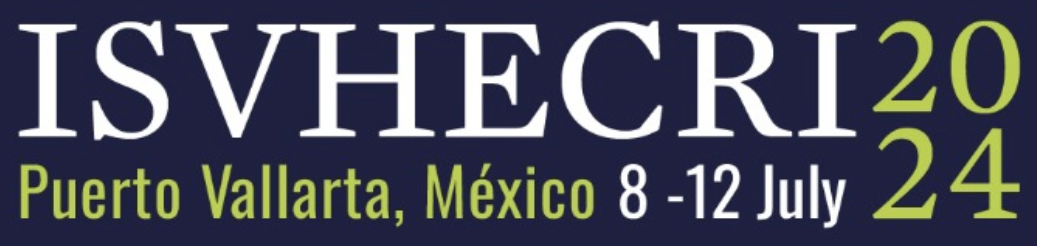}
  \end{minipage}
  &
  \begin{minipage}{0.55\textwidth}
    \begin{center} \hspace{5pt}
    {\it 22nd International Symposium on Very High \\Energy Cosmic Ray Interactions (ISVHECRI 2024)} \\
    {\it Puerto Vallarta, Mexico, 8-12 July 2024} \\
    \doi{10.21468/SciPostPhysProc.?}\\
    \end{center}
  \end{minipage}
\end{tabular}
}
\end{center}

\section*{\color{scipostdeepblue}{Abstract}}
\textbf{\boldmath{
The Pierre Auger Observatory has revealed a significant challenge in air shower physics: a discrepancy between the simulated and observed muon content in cosmic-ray interactions, known as the `Muon Puzzle`. This issue stems from a lack of understanding of high-energy hadronic interactions. Current state-of-the-art hadronic interaction models fall short, underscoring the need for improvements. In this contribution, we explore the integration of the \pythia 8 hadronic interaction model into air shower simulations. 
While \pythia 8 is primarily used in Large Hadron Collider experiments, recent advancements in its Angantyr module show promise in better describing hadron-nucleus interactions, making it a valuable tool for addressing the Muon Puzzle.
}}

\vspace{\baselineskip}

\noindent\textcolor{white!90!black}{%
\fbox{\parbox{0.975\linewidth}{%
\textcolor{white!40!black}{\begin{tabular}{lr}%
  \begin{minipage}{0.6\textwidth}%
    {\small Copyright attribution to authors. \newline
    This work is a submission to SciPost Phys. Proc. \newline
    License information to appear upon publication. \newline
    Publication information to appear upon publication.}
  \end{minipage} & \begin{minipage}{0.4\textwidth}
    {\small Received Date \newline Accepted Date \newline Published Date}%
  \end{minipage}
\end{tabular}}
}}
}


\vspace{10pt}
\noindent\rule{\textwidth}{1pt}

\section{Introduction}
\label{sec:intro}

Air showers are phenomena originating from the interaction of particles, such as cosmic rays, with the Earth's atmosphere. This primary interaction initiates cascades of secondary particles, divided hadronic, electromagnetic, and muonic components. The Pierre Auger Observatory~\cite{PierreAuger:2015eyc} studies these particles using Cherenkov light in water tanks and fluorescence light from the deexcitation of N$_2$ molecules in the atmosphere. Air shower simulations are crucial to translate these detector signals into meaningful physical quantities. The electromagnetic shower profile driven by neutral pions decaying into photons, involves pair production and bremsstrahlung from electrons, from which the maximum shower depth X$_\text{max}$ can be inferred. While produced in hadronic interactions and decays, muons are of importance to study due to their deep penetration and distinct detector signals, with the muon number N$_\mu$ serving as a key observable for determining cosmic ray mass composition. 

Current status regarding the study of muons from air shower data shows a significant muon deficit in air shower simulations with respect to measurements from the Pierre Auger Observatory~\cite{PierreAuger:2014ucz, PierreAuger:2020gxz, Soldin:2021wyv}, from the TeV scale increasing with energy. Hadronic interaction models must be extrapolated to higher energies for air showers to be simulated correctly, as their phenomenological counterparts remain unvalidated by collider data, and the discrepancy rooted in secondary particle production. While the forward phase space is vital for air shower studies, exploration has been limited to fixed-target experiments and forward detectors thus far.

Several studies already took place to gain insights into the Muon puzzle: from ad hoc modification of cross-section, multiplicity, and elasticity of hadronic interactions model~\cite{Ulrich:2010rg, Ebr:2023nkf} or by altering directly particle production~\cite{Riehn:2024prp}, to perform a multi-parameter fit of model predictions against Pierre Auger data~\cite{Vicha:2022zvv, Vicha:2023jup}. Yet the puzzle remains unsolved. To explore the potential of its origins further, this work introduces another hadronic interaction model into the landscape of air showers, for which all above-mentioned study can be applied to. 

\section{\pythia 8}

\pythia 8~\cite{Bierlich:2022pfr} is an event generator that describes interactions between protons, electrons, photons, and heavy nuclei in high-energy physics collisions. The nuclear interactions are handled by the Angantyr model~\cite{Bierlich:2018xfw, Bierlich:2021poz} in \pythia 8, bridging the gap between high energy and heavy ion phenomenology. A variety of physics topics are covered by \pythia 8, including total and partial cross sections, hard and soft interactions, parton distributions, initial- and final-state parton showers, multiparton interactions, hadronization/fragmentation and decays. 

\begin{figure}[h!]
    \centering
    \includegraphics[width=0.55\linewidth]{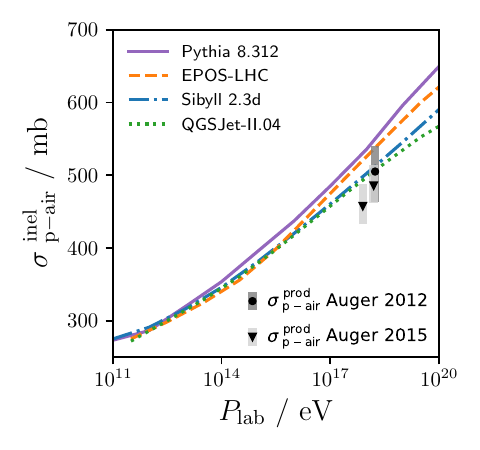}
    \caption{\small Inelastic cross-sections distributions (solid lines) as a function of the momentum in the laboratory frame of the projectile for proton-air collisions, assuming the composition of air as 80\% nitrogen-14 to 20\% oxygen-16 mix. Production cross-sections$^{1}$ (markers) for proton-air collisions from \cite{PierreAuger:2012egl, Ulrich:2015yoo}.}
    \label{fig:xsec_p-air}
\end{figure}

Comparisons of inelastic cross-sections between \pythia to other hadronic interaction models~\cite{Pierog:2013ria, Pierog:2023ahq, Riehn:2019jet, Ostapchenko:2010vb} for a proton-air collisions are displayed in Fig.~\ref{fig:xsec_p-air}. 
The description offered by \pythia for proton-air interactions is in agreement within the most commonly used models. Removing the quasi-elastic contribution from the inelastic cross-section of \pythia may yield cross-section values that are more compatible with Auger data. \footnotetext[1]{Production and inelastic cross-sections are not directly comparable, as $\sigma^\mathrm{inel} = \sigma^\mathrm{prod} + \sigma^\mathrm{quasi-elastic}$. The inelastic cross-section distributions can be considered as upper limits to the production cross-section ones.}

\begin{figure}
    \centering
    \includegraphics[width=0.85\linewidth]{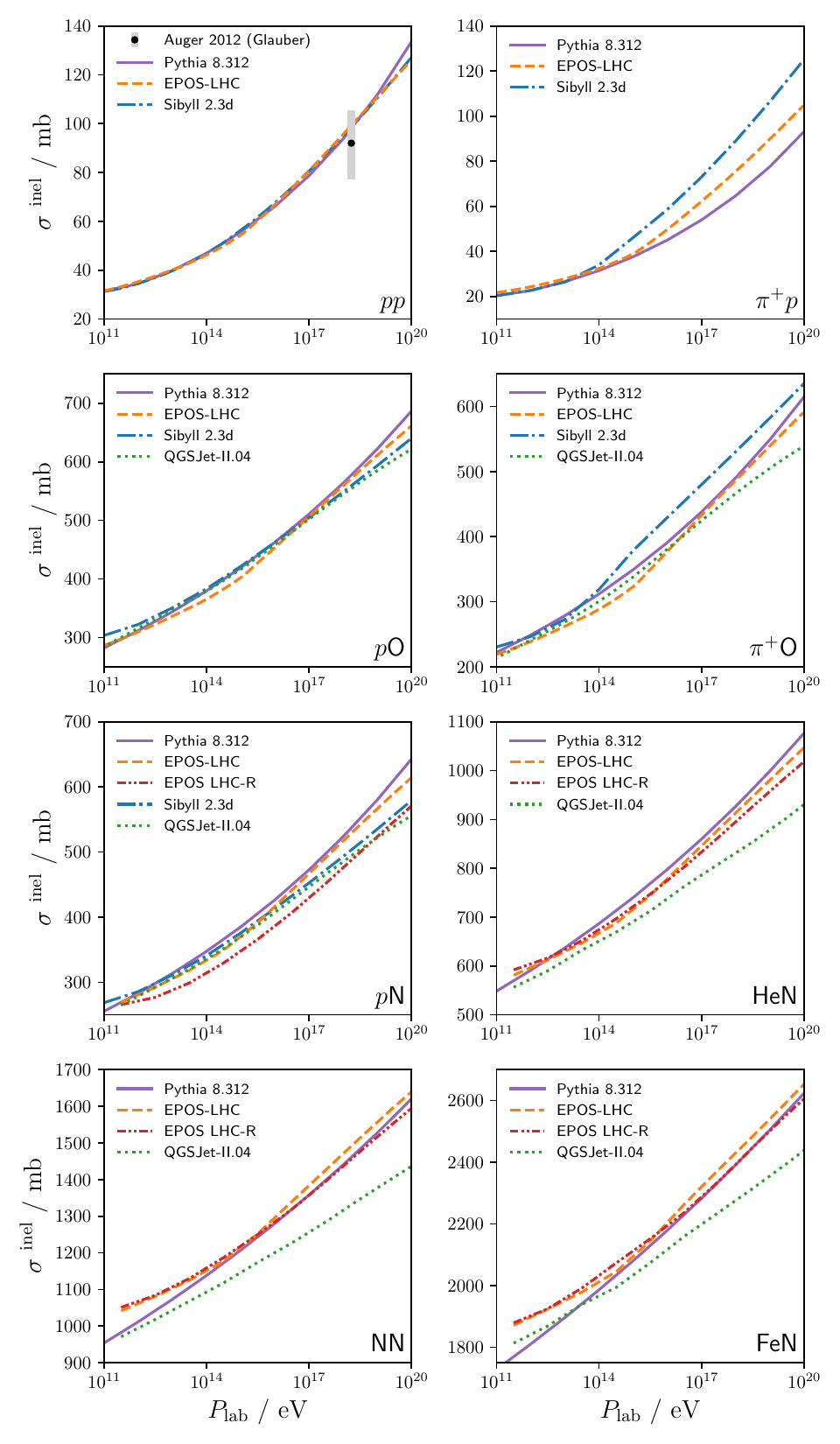}
    \caption{\small Inelastic cross-sections as a function of the momentum in the laboratory frame of the projectile for several collision systems. Top row: pp and $\pi^{+}$p collisions; Second row from top: p-$^{16}$O and $\pi^{+}$-$^{16}$O collisions; Third row from top: p-$^{14}$N and $^{4}$He-$^{14}$N collisions. Bottom row: $^{14}$N-$^{14}$N and $^{56}$Fe-$^{14}$N collisions. Results from EPOS LHC-R are preliminary.}
    \label{fig:xsec}
\end{figure}

Further comparisons is done with other collision systems displayed in Fig.~\ref{fig:xsec}, including additionally a preliminary version of EPOS LHC-R. Although results from \pythia are consistent with most models, as well as experimental results in the case proton-proton collisions, it predicts a steeper slope in the highest energy region compared to other models visible for $p$N, $p$O, HeN and NN interactions present in air showers. This feature would result in larger hadronic cross-sections at energies comparable to ultra-high-energy cosmic-ray primaries. It would impact the muon production as it introduces shift of depth of the shower maximum hinting at an earlier development of the shower, resulting in further muons detected at ground, and therefore reducing the gap between simulations and experimental data~\cite{PierreAuger:2014ucz}. Further investigation of the fluctuations observed for the $\pi^{+}$O and FeN collision systems needs to take place.

Recent updates to the Angantyr model are making its usage attractive for air shower studies. Efforts are ongoing to bring \pythia 8, with Angantyr, in the landscape of shower simulation codes such as \corsika 8~\cite{Engel:2018akg} and MCEq~\cite{Fedynitch:2015zma} to study shower development and lepton fluxes.

While \pythia 8  may be run independently and has internal tools to do basic analysis, there are interfaces for input and output to other programs such as \rivet. \rivet~\cite{Bierlich:2019rhm} was created to support the development, validation, and tuning of event generators. A \rivet plugin is a C++ program that preserves the original analysis logic in a reproducible form. It stores measurements in human-readable YODA format~\cite{Buckley:2023xqh}, which can be manually written or imported from HEPData~\cite{Maguire:2017ypu}. The plugin's C++ code defines how to transform event generator output to match measurements, including particle selection, event filtering, and histogram binning. \pythia 8 can feed the raw events directly into \rivet via \texttt{Pythia8Rivet}, while other models use the HepMC event record~\cite{Dobbs:2001ck} as an intermediate format.

\begin{figure}[h]
    \centering
        \begin{overpic}[width=0.6\linewidth, grid=false, tics=10]{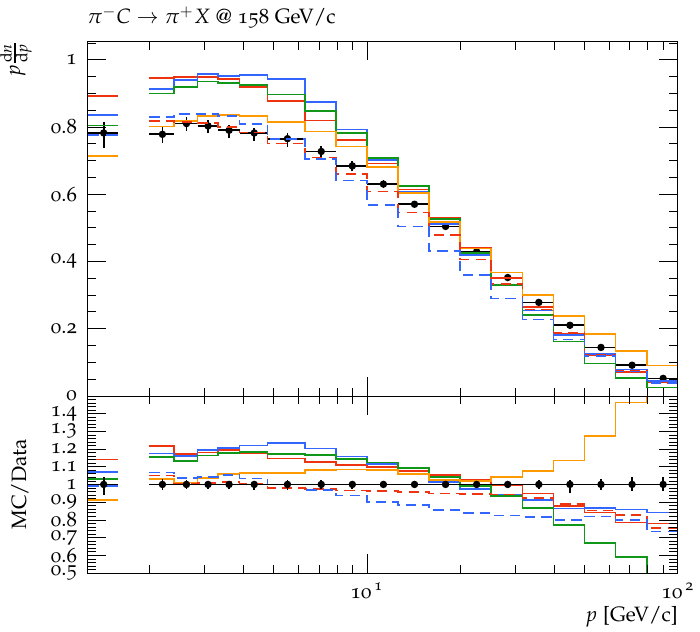}
            \put(65,64.5){ 
                \begin{scriptsize}
                \begin{tikzpicture}
                    \tikzstyle{NA61}=[black, line width=1pt]
                    \tikzstyle{EPOS-LHC}=[RRed, line width=1pt]
                    \tikzstyle{Sibyll-2.3d}=[RBlue, line width=1pt]
                    \tikzstyle{Pythia}=[RGreen, line width=1pt]
                    \tikzstyle{QGSJet}=[ROrange, line width=1pt]
    
                    \draw[NA61] (0,0) -- (0.5,0);
                    \fill[black] (0.25,0) circle (0.07);
                    \draw[black, line width=0.5pt] (0.25,-0.2) -- (0.25,0.2);
                    \node[right, xshift=0.15cm] at (0.5,0) {NA61/SHINE};
    
                    \draw[EPOS-LHC] (0,-0.3) -- (0.5,-0.3);
                    \node[right, xshift=0.15cm] at (0.5,-0.3) {EPOS-LHC};
                
                    \draw[Sibyll-2.3d] (0,-0.6) -- (0.5,-0.6);
                    \node[right, xshift=0.15cm] at (0.5,-0.6) {Sibyll 2.3d};
                
                    \draw[Pythia] (0,-0.9) -- (0.5,-0.9);
                    \node[right, xshift=0.15cm] at (0.5,-0.9) {Pythia 8.312};
                
                    \draw[QGSJet] (0,-1.2) -- (0.5,-1.2);
                    \node[right, xshift=0.15cm] at (0.5,-1.2) {QGSJet-II.04};
                \end{tikzpicture}
                \end{scriptsize}
            }
            \put(13,35){ 
            \begin{scriptsize}
            \begin{tikzpicture}
                \tikzstyle{EPOS-199}=[RRed, line width=1pt, dashed]
                \tikzstyle{Sibyll-2.1}=[RBlue, line width=1pt, dashed]

                \draw[EPOS-199] (0,0) -- (0.5,0);
                \node[right, xshift=0.15cm] at (0.5,0) {EPOS 1.99};
            
                \draw[Sibyll-2.1] (0,-0.3) -- (0.5,-0.3);
                \node[right, xshift=0.15cm] at (0.5,-0.3) {Sibyll 2.1};
            \end{tikzpicture}
            \end{scriptsize}
        }
        \end{overpic}
        \caption{Simulations of the $\pi^{+}$ production in $\pi^{-}$C collisions at $158$ GeV as a function of the outgoing $\pi^{+}$ momenta, for several hadronic models against NA61/SHINE data~\cite{NA61SHINE:2022tiz}. Dashed: pre-LHC tuned models; Solid: LHC tuned models.}
        \label{tab:rivet_plot}
    \end{figure}
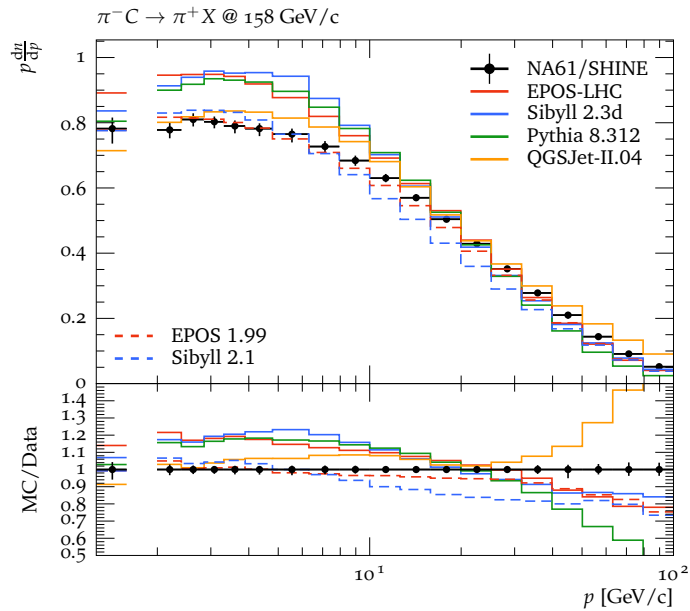

\paragraph*{\corsika 8 + \pythia 8}
A previous implementation of \pythia 8 in \corsika 8 was done for version 8.3.07 using the \texttt{PythiaCascade} class for hadron-ion collision systems~\cite{Reininghaus:2023ctx, Reininghaus:2022gnr}. This work follows in the footsteps of previous study, but now focusing on the Angantyr model in \pythia 8.312 and its opportunities.

The current implementation of \pythia in its 8.312 version, alongside with the Angantyr model, in \corsika 8 is at a preliminary stage. After initializing the shower input setting, \corsika calls either \pythia or Angantyr depending on the collision system at hand (pp or hA/AA). The latest update brings energy and beam switching features into play, allowing a single \pythia instance to run and perform any collisions for a given projectile, target and center-of-mass energy on an event-by-event basis. The inelastic cross-section is interpolated from pre-generated tables, and secondary final state particles are obtained for each collision system. Tracking occurs internally within \corsika, storing all shower events in a particle stack, resulting in \corsika output files offering insights into shower profiles and ground-level particles among other information.

All tests run with version 8.312 were targeted at validating proton-proton collisions between \pythia wherin \corsika 8 and standalone \pythia simulations, as well as checking proton primary air showers generation. Before running air shower simulations with nuclear primaries, several issues must be resolved: the long initialization time required for Angantyr, some gaps in the tabulated cross-sections for nuclear projectiles and targets, and the handling of the nuclear remnants. Optimizing the usage of reuse files within \pythia may cut down initialization time of Angantyr, while adopting a semi-superposition model (similarly done for Sibyll 2.3d) could bridge the gaps in the cross-section tables.

\section{Tuning of \pythia 8}

Tuning of event generators can be done manually by adjusting parameters until generator predictions match experimental results in control plots. While manual tuning is a lengthy process, a more robust approach is employing optimization algorithms which is analogous to a multidimensional fit. However tuning these models remains a complex task that requires significant domain expertise and cooperation between different working groups, from providing sensible inputs during parameter selection, as well as observable weights, to the design of validation tests and their subsequent interpretation. For the standard Monash tune~\cite{Skands:2014pea}, \pythia 8 is tuned to data from e$^+$e$^-$ annihilation and pp/p$\overline{\text{p}}$ collisions, but not using collision systems common in air showers (proton-air, pion-air, kaon-air).

\begin{figure}[h!]
    \centering
    \includegraphics[width=0.8\linewidth]{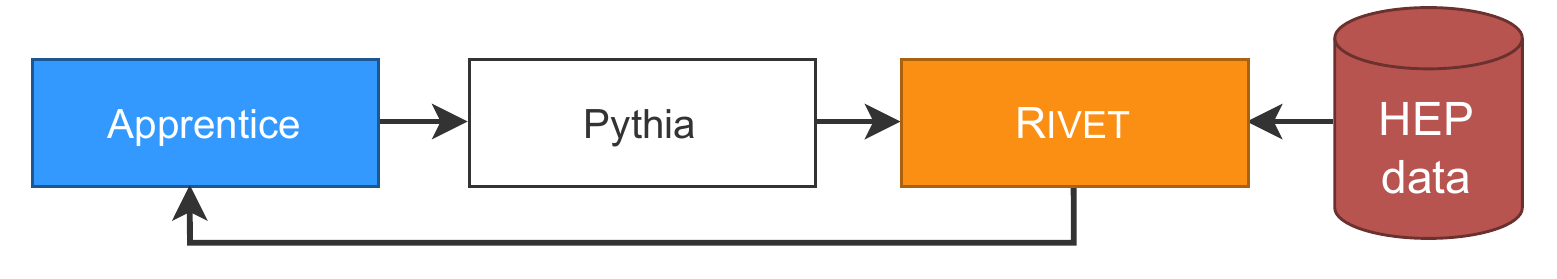}
    \caption{Visualization of the information flow employed in the tuning of \pythia.}
    \label{fig:classic_tune}
\end{figure}

This tuning endeavour starts with accelerator data to tune \pythia 8 against, along with the parameters of interest that adjust \pythia's physics settings. By probing the selected parameter space, Monte Carlo samples are generated and compared with experimental datasets using \rivet. A polynomial approximation of the Monte Carlo data is then computed, increasing in accuracy as both the polynomial order and sample size grow. This fit is provided as input to the tuning software \textsc{Apprentice}~\cite{Krishnamoorthy:2021nwv}, which applies various minimizing algorithms to determine the best fit for the chosen \pythia parameters. Once tuned, \pythia is run through \rivet to produce cross-check plots against several experimental results. The tuned version can then be incorporated into both the air shower simulation code \corsika 8 and the cascade equation solver MCEq, allowing an analysis of the impact of the tuned parameters on muon production when compared to a default \pythia description.

\paragraph*{\pythia 8 Wuppertal tune}
This work discusses a tune of \pythia 8 to better describe forward hadron production, total and inelastic cross sections for different initial collision systems measured in fixed-target experiments at the SPS and Fermilab. These accelerator data summarized in Table~\ref{tab:rivet_table}, span centre-of-mass energies of $\sqrt{s} \sim 6$ to $20$ GeV, much lower than the LHCf measurements targeted in the forward \pythia tune~\cite{Fieg:2023kld}. As of now, only few \rivet plugins relevant for the Wuppertal tune are part of the \rivet analysis library while most are in the writing, and will be used within the \textsc{Apprentice} framework to tune \pythia 8, as displayed in Fig.~\ref{fig:classic_tune}. On a step-by-step basis, the tune parameters will be fixed for each collision system, beginning with pp, then $\pi^{\pm}$p/K$^{\pm}$p, followed by pC and ending with $\pi$C which mimic proton-air and pion-air collisions. The tune will focus on similar parameters as the forward \pythia tune, namely parameters related to the beam remnants and the popcorn mechanism. 

\begin{table}[h!]
    \centering
    \begin{tabular}{c c c l c c}
        \hline
        Paper  & HEPData  & \rivet  & Observable  & $p_{\rm{beam}}$ (GeV/c)  & Collision system \\ \hline \hline
        \cite{Carroll:1975xf}         & \href{https://www.hepdata.net/record/ins98502}{$\checkmark$} &  & \raisebox{-.25\height}{$\sigma^\mathrm{tot}$}
        & [23 - 280]                  & $\pi^\pm$p, K$^\pm$p, pp, $\overline{\rm{p}}$p \\
        \cite{Carroll:1978vq}         & \href{https://www.hepdata.net/record/ins132765}{$\checkmark$} &  & \raisebox{-.25\height}{$\sigma^\mathrm{tot}$}
        & [200 - 370]                 & $\pi^\pm$p, K$^\pm$p, pp, $\overline{\rm{p}}$p \\
        \cite{Carroll:1978hc}         & \href{https://inspirehep.net/literature/132133}{$\checkmark$} &  & \raisebox{-.25\height}{${\sigma^\mathrm{absorption}}$}
        & [60 - 280]                  & $\pi^\pm$p, K$^\pm$p, pp, $\overline{\rm{p}}$p \\
        \cite{Burq:1982ja}            & \href{https://www.hepdata.net/record/ins182455}{$\checkmark$} &  & elastic $\frac{\text{d}\sigma}{\text{d}t}$
        & [30 - 345]                  & $\pi^{-}$p, pp \\
        \cite{NA22:1987lmr}           & \href{https://www.hepdata.net/record/ins246909}{$\checkmark$} &  & elastic $\frac{\text{d}\sigma}{\text{d}t}$ 
        & 250                         & $\pi^{+}$p, K$^{+}$p, pp \\
        \cite{EHSNA22:1988fqa}        & \href{https://www.hepdata.net/record/ins265504}{$\checkmark$} & \href{https://rivet.hepforge.org/analyses/EHS_1988_I265504}{$\checkmark$} & charged particle production
        & 250                         & $\pi^{+}$p, K$^{+}$p, pp \\
        \cite{EHSNA22:1990vem}        & \href{https://www.hepdata.net/record/ins301243}{$\checkmark$} &  & $\rho^0$, $\rho^+$, $\omega$ production 
        & 250                         & $\pi^{+}$p \\
        \cite{EHSNA22:1991dhh}        & \href{https://www.hepdata.net/record/ins322980}{$\checkmark$} &  & $\pi^0$, $\eta$ production 
        & 250                         & $\pi^{+}$p, K$^{+}$p \\
        \cite{NA49:2005qor}           & \href{https://www.hepdata.net/record/ins694016}{$\checkmark$} & \href{https://rivet.hepforge.org/analyses/NA49_2006_I694016.html}{$\checkmark$} & $\pi^\pm$ production 
        & 158                         & pp \\
        \cite{NA49:2009brx}           &  & \href{https://rivet.hepforge.org/analyses/NA49_2009_I818217.html}{$\checkmark$}  & p, $\overline{\text{p}}$, n production
        & 158                         & pp \\
        \cite{NA61SHINE:2017fne}      & \href{https://www.hepdata.net/record/ins1598505}{$\checkmark$} &  & $\pi^\pm$, K$^\pm$, p, $\overline{\text{p}}$ production 
        & [20 - 158]                  & pp \\
        \cite{NA61SHINE:2019aip}      & \href{https://www.hepdata.net/record/ins1753094}{$\checkmark$} &  & \raisebox{-.25\height}{$\sigma^\mathrm{prod}$, $\sigma^\mathrm{inel}$} 
        & 60, 120                     & pC, pBe, pAl \\
        \cite{NA61SHINE:2017vqs}      & & & $\rho^0$, $\omega$, K$^{{0}^{*}}$ \hspace{-0.15cm} production 
        & 158, 350 & $\pi^{-}$C  \\
        \cite{NA61SHINE:2022tiz}      &  &  & $\pi^\pm$, K$^\pm$, p, $\overline{\text{p}}$ production 
        & 158, 350                    & $\pi^{-}$C \\ \hline
    \end{tabular}
    \caption{Status of the \textsc{Rivet} plugins and HEPData entries of interest for the \pythia~8 Wuppertal tune.}
    \label{tab:rivet_table}
\end{table}

In the future, data from proton-oxygen runs at the LHC, will be of great interest for air shower studies. They could be included in an extension of this tune to add another reference point at the TeV scale which is extremely relevant for solving the Muon Puzzle~\cite{Albrecht:2021cxw}.

\paragraph*{Global tune for extensive air showers} 
A tune including, in addition to accelerator datasets, air shower observables is an emerging idea among our community. Such air shower observables would be muonic shower properties (N$_\mu$(E), $\sigma$(N$_\mu$, R$_\mu$, ...), electromagnetic shower properties (N$_\text{e}$ , X$_\text{max}$, $\sigma$(X$_\text{max}$), ...), cross-sections inferred from X$_\text{max}$ distribution tail, mass composition of cosmic rays. This has been extensively discussed at the workshop on the tuning of hadronic interaction models\footnote{\href{https://indico.uni-wuppertal.de/event/284/}{Workshop's indico}}.

\section{Conclusion}

The integration of \pythia 8 into air shower simulations represents a promising step forward in addressing the long-standing `Muon Puzzle`. With the Angantyr model, this work bridges the gap between high-energy particle physics and cosmic-ray studies. Preliminary results highlight the potential for improving muon predictions, although further tuning and validation with accelerator data remain essential. Future efforts will focus on refining the cross-section tables fed to \corsika 8, and addressing runtime initialization and nuclear remnants challenges related the Angantyr model. Alongside, the development of \rivet plugins tailored to the \pythia 8 Wuppertal tune will continue. These improvements will ultimately provide a clearer understanding of the discrepancies in muon content, enhancing our comprehension of high-energy cosmic-ray events.

\section*{Acknowledgements}
\paragraph{Funding information}
This work is supported by the German Research Foundation (DFG, Deutsche Forschungsgemeinschaft) via the Collaborative Research Center SFB1491: Cosmic Interacting Matters – from Source to Signal (F4) -- project no. 445052434.

\bibliography{ISVHECRI_2024.bib}

\end{document}